%% file: short.tex
\documentclass{easychair}

\usepackage{amsthm}
\usepackage{xspace}
\usepackage[utf8]{inputenc}
\usepackage{paralist}
\usepackage{bold-extra}
\usepackage{textcomp}
\usepackage{xcolor}
\usepackage{listings}
\usepackage{isabelle,isabellesym}
\usepackage{hyperref}
\usepackage{microtype}

\definecolor{dblue}{rgb}{0.2,0.2,0.7}%
\hypersetup{%
  colorlinks=true,%
  pdfborder={0 0 0},%
  linkcolor=dblue,%
  citecolor=dblue,%
  urlcolor=dblue,%
}

\makeatletter
\newcommand*\verbfont{\normalfont\ttfamily
  \@noligs}
\makeatother
\lstnewenvironment{tty}{\lstset{
  basicstyle=\verbfont,%
  flexiblecolumns=false,%
  upquote=true,%
  basewidth={0.5em,0.45em},%
  }}{}
\newcommand{\hs}{\lstinline[style=Haskell]}
\lstnewenvironment{haskell}{\lstset{style=Haskell}}{}
\lstdefinelanguage{Haskell}{%
  morekeywords=,%
  morekeywords=[2]{%
    case,class,data,default,deriving,do,else,%
    foreign,if,import,in,infix,infixl,infixr,instance,%
    let,module,newtype,of,then,type,where,\_,%
  },%
  morecomment=[l]--,%
  morecomment=[n]{\{-}{-\}},%
  sensitive=true,%
  morestring=[b]",
  showstringspaces=false,%
}[keywords,comments,strings]%

\lstdefinestyle{Verbatim}{%
  upquote=true,%
  xleftmargin=2.5ex,%
  xrightmargin=2.5ex,%
  keepspaces=True,%
  basicstyle=\upshape\ttfamily,%
  flexiblecolumns=false,%
  basewidth={0.5em,0.45em},%
  aboveskip=\smallskipamount,%
  belowskip=\smallskipamount,%
}
\lstdefinestyle{Haskell}{%
  style=Verbatim,%
  language=Haskell,%
  identifierstyle=,%
  keywordstyle=,%
  keywordstyle={[2]\bfseries},%
  commentstyle=,%
  stringstyle=,%
}
\newenvironment{isamargin}{%
  \begin{list}{}{%
    \setlength{\topsep}{\smallskipamount}%
    \setlength{\leftmargin}{2.5ex}%
    \setlength{\rightmargin}{2.5ex}%
    \setlength{\listparindent}{\parindent}%
    \setlength{\itemindent}{0pt}%
    \setlength{\parsep}{\parskip}%
  }%
  \item[]}{\end{list}}
\newcommand{\DefineSnippet}[2]{%
  \expandafter\newcommand\csname snippet--#1\endcsname{%
    \begin{isamargin}
    \begin{isabelle}
    #2
    \end{isabelle}
    \end{isamargin}
  }}
\newcommand{\Snippet}[1]{\csname snippet--#1\endcsname}
\isabellestyle{it}
\renewcommand\isastyletxt{\isastyletext}
\renewcommand\isastyle{\isastyleminor}
\renewcommand{\isachardoublequoteopen}{``{}}
\renewcommand{\isachardoublequoteclose}{''{}}

\renewcommand{\isacharunderscore}{\_}

\newcommand\BC[1]{\item \textbf{Base case} (#1).}
\newcommand\SC[1]{\item \textbf{Step case} (#1).}
\newcommand\holcfpversion{0.1\xspace}
\newcommand\hlintversion{1.8.46\xspace}
\newcommand\hlint{\texttt{HLint}\xspace}
\newcommand\file[1]{\nolinkurl{#1}}
\newcommand\ignore[1]{}

\title{Certified \hlint{}s with Isabelle/HOLCF-Prelude}
\titlerunning{Certified \hlint{}s with Isabelle/HOLCF-Prelude}

\author{
  Joachim Breitner\inst{1}\thanks{Supported by the Deutsche Telekom Stiftung.}
\and
  Brian Huffman\inst{2}
\and
  Neil Mitchell\inst{3}
\and
  Christian Sternagel\inst{4}\thanks{Supported by the Austrian Science Fund (FWF): J3202.}
}

\institute{
  Karlsruhe Institute of Technology,
  \email{breitner@kit.edu}
\and
  Galois, Inc.,
  \email{huffman@galois.com}
\and
  \email{ndmitchell@gmail.com}
\and
  JAIST,
  \email{c-sterna@jaist.ac.jp}
}

\authorrunning{J. Breitner, B. Huffman, N. Mitchell and C. Sternagel}

\input{snippets}

\begin{document}
\maketitle

\begin{abstract}
We present the HOLCF-Prelude, a formalization of a large part of Haskell's
standard prelude in Isabelle/HOLCF. Applying this formalization to the hints
suggested by \hlint allows us to certify them formally. 
\end{abstract}

In pure functional languages such as Haskell, equational reasoning is a valuable
tool for refactoring, to improve both efficiency and aesthetics. For example, an
experienced programmer would replace
\hs{reverse ".txt" `isPrefixOf` reverse filename}
with the more readable (and more efficient)
\hs{".txt" `isSuffixOf` filename}.
In this paper we call such a replacement a \emph{rewrite}. We only
want to apply rewrites that are valid and thus some natural questions arise:  \emph{Is
the original expression equivalent to the replaced expression?} With a language
like Haskell, this entails the question: \emph{What about when
infinite or undefined values are involved?}

To highlight some of the issues, consider another example.  Assuming the
definition
\begin{haskell}
reverse []     = []
reverse (x:xs) = reverse xs ++ [x]
\end{haskell}
can we safely apply the following rewrite?
\begin{equation}
\text{\hs{reverse (reverse xs)}} = \text{\hs{xs}} \tag{$\star$}\label{revrev}
\end{equation}
Let us try to prove \eqref{revrev} by induction:

\begin{compactitem}
\BC{\hs{xs = []}}
Just apply the definition of \hs{reverse}.

\SC{\hs{xs = y:ys}}
We have:

\noindent\begin{tabular}{l@{\quad}l@{\quad}l}
\multicolumn{3}{l}{\hs{reverse (reverse (y:ys))}}\\
&= \hs-reverse (reverse ys ++ [y])- & (by definition of \hs-reverse-)\\
&= \hs-reverse [y] ++ reverse (reverse ys)- & (using an auxiliary lemma)\\
&= \hs-reverse [y] ++ ys- & (by induction hypothesis)\\
&= \hs-y:ys-
\end{tabular}
\end{compactitem}
Such fast-and-loose reasoning \cite{fastandloose} is oftentimes useful, but may
fail for lazy languages: The above rewrite is neither valid for infinite
\hs{xs}, nor when \hs{xs} contains undefined values on the spine.
In addition to the above cases, we should have considered the undefined input
$\bot$ (pronounced \emph{bottom}) and made sure that the desired property is
\emph{admissible} for our setting.\footnote{See \cite{Huffman2011} for a formal
definition of admissibility.}
These extra requirements can be tricky to follow, so automated assistance would
be welcome.

Such assistance is available using higher-order logic for computable functions
(HOLCF, \cite{Huffman2011}).  HOLCF is based on the higher-order logic (HOL)
instance of the proof assistant Isabelle \cite{Isabelle} that provides
functions, recursive definitions, (data)
types, type classes, etc.; and constitutes a domain-theoretic framework that
allows us to generate types in HOL that match the denotation of types in
Haskell, i.e., with possibly infinite values and explicit bottom values. With these
pieces we can define functions using Haskell-like pattern matching with
call-by-need semantics which can handle both laziness and
infinite data structures. The definition of \hs{reverse} carries over quite
naturally:
\Snippet{reverse-def}
Note that since Isabelle/HOL's default function type represents total functions,
there is the special notation `$\cdot$' for application of Haskell-like (i.e.,
continuous) functions.

In Isabelle/HOLCF every domain is equipped with a partial order $\sqsubseteq$
whose least element is $\bot$. We say \emph{$s$ is less defined than $t$}, whenever $s \sqsubset t$.
To formalize our proof about \hs{reverse} in Isabelle/HOLCF we switch from
equality to $\sqsubseteq$. First we show how reverse ``distributes'' over
list-append:
\Snippet{reverse-append-below}
Then we obtain the desired lemma:
\Snippet{reverse-reverse-below}
In both cases, the proofs just require induction followed by equational
reasoning (\emph{simplification} in Isabelle parlance), where all cases except
for the step-case are trivial (i.e., solved automatically).


In order to make Isabelle/HOLCF more useful for the verification of Haskell
programs, we have started to formalize some Haskell standard modules \cite{prelude}.
The result is the ongoing open source project
Isabelle/HOLCF-Prelude\footnote{\url{http://sourceforge.net/p/holcf-prelude/}}
(HOLCF-Prelude for short). Contributions are most welcome and you
may obtain the corresponding mercurial repository via
\begin{lstlisting}[style=Verbatim]
  hg clone http://hg.code.sf.net/p/holcf-prelude/code holcf-prelude
\end{lstlisting}
As of version
\holcfpversion, it contains theories about booleans, the \hs{Maybe} type, integers,
tuples, lists, functions on those types, as well as the type classes \hs{Eq} and
\hs{Ord}.

The tool \hlint\footnote{\url{http://community.haskell.org/~ndm/hlint/}}
(version \hlintversion) suggests improvements to Haskell code. Example
suggestions include using more appropriate functions, eliminating redundant
language extension pragmas and avoiding excessive bracketing. Many suggestions
are rewrites, which are called \emph{hints}.
In hints, all single-letter variables are treated as free variables and the
expression they match on the left-hand side is substituted on the right-hand
side.
\hlint also allows a severity level (like \hs{error} and \hs{warn}ing) and
notes to be associated with each hint.  Notes are presented to the user along
with the hint.  Most hints represent equalities and have no note, but some are
only true in certain circumstances.
The majority of \hlint's hints are ``obvious,'' but there are many of them, contributed
by a large number of people. Manually checking the hints is error-prone, and
several bugs have been reported by end-users (quite possibly after modifying
their code in response to a hint).

The certification of these hints is our first application of the HOLCF-Prelude.
Consider the hint (of severity level \hs{warn})
\begin{haskell}
warn = reverse (reverse x) ==> x where note = IncreasesLaziness
\end{haskell}
which says that you should replace
\hs{reverse (reverse x)} in your code by \hs{x}, but also notes that such a
replacement will possibly increase the laziness of the program,
meaning that there may be situations in which the original code crashes or does not terminate,
while it will not do so after applying the hint.
If we also have the file \file{test.hs} containing
\begin{haskell}
output xs = print (reverse (reverse (sort xs)))
\end{haskell}
and run \hlint on the file, it will respond with:
\begin{tty}
  test.hs:1:20: Warning: Use alternative
  Found:
    reverse (reverse (sort xs))
  Why not:
    sort xs
  Note: increases laziness
\end{tty}
That is, \hlint suggests a rewrite, and warns the user that the resulting
expression is lazier than the original one, so if strictness was the purpose of
using \hs{reverse} the replacement may not be desirable.\footnote{Anyone wanting
a spine-strict list would be better off using the more efficient
\hs{length x `seq` x} pattern.}

In order to facilitate the formal verification of such hints, we have modified
\hlint to generate Isabelle/HOLCF lemmas. We can do this for the above hint by running
\begin{tty}
  hlint \
    --with='warn = reverse (reverse x) ==> x where note = IncreasesLaziness' \
    --proof=/dev/null --report
\end{tty}
which generates a file \file{report.txt} containing the above hint in Isabelle
notation:
\begin{tty}
  reverse\<cdot>(reverse\<cdot>x) \<sqsubseteq> x
\end{tty}
This we turn into the lemma (as shown in an Isabelle UI, such as Isabelle/jEdit or Isabelle/Proof\,General)
\Snippet{reverse-reverse-below-goal}
that we have proven above.

Proofs for many of \hlint's default hints are already part of the HOLCF-Prelude.
During our formalization we uncovered three previously unknown errors in \hlint (and many
missing annotations):
\begin{compactitem}
\item
The hint \hs{take (length x - 1) x ==> init x} introduces a crash on the empty list
(and thus was removed from \hlint's database).

\item
The hint \hs{head (drop n x) ==> x !! n} is only true if the index is
non-negative (and thus was modified to include this condition).

\item
The hint \hs{take i s == t ==> (i == length t) && (t `isPrefixOf` s)} was
found to be erroneous (and thus was removed from \hlint's database).
\end{compactitem}

Before starting this formalization effort a handful of hints had laziness
annotations, but they were not intended to be complete. With the new scheme we
know (for the proved hints) that we did not miss any annotations.

\paragraph{Known Issues.}

We haven proven some \hlint hints correct, but while we can have confidence that the hint itself is correct, there are still ways the user can end up with incorrect rewrites.

The validity of any rewrite depends on the definitions of the functions
involved. The Haskell standard \cite{haskell-report} contains implementation
suggestions for many functions, mostly aiming for simplicity and elegance, and
we follow these definitions. In real compilers, for example GHC, the actual
implementation is often somewhat different (look out for
\file{USE_REPORT_PRELUDE} in the sources). In many cases, the definition is
believed to be equivalent -- for example \hs{splitAt} in the GHC sources is
performed with a single traversal of the list and unboxed \hs{Int#} values, while the standard defines \hs{splitAt} in terms of \hs{take} and \hs{drop}. In other cases the definition is only morally equivalent -- for example \hs{elem} is only equivalent for commutative definitions of \hs{Eq}.

Another issue are type classes. Currently HOLCF-Prelude supports \hs{Eq} and \hs{Ord}. While it might be tempting to assume that the former implements an equivalence relation, such properties are not enforced by Haskell. What constitutes a valid instance of \hs{Eq}? For maximal flexibility, we distinguish several cases in our formalization, among them: \isa{Eq} (just syntactic, i.e., functions \isa{eq} and \isa{neq} are available for the type, where the default implementation for \isa{neq} is assumed), \isa{Eq\_sym} (assuming \isa{eq} is strict and symmetric) and \isa{Eq\_equiv} (extends \isa{Eq\_sym} to an equivalence relation).  The question arises, when \hlint suggests a rewrite involving \isa{Eq}, what kind of properties may we assume? Currently we require annotations on all hints requiring properties of \hs{Eq}, translate them to \isa{Eq\_sym} for the proofs, and display notes to the user when suggesting such replacements.

Another potential source of errors is in \hlint itself. While the hint may be
true, \hlint has complicated unification routines tuned for performance, and
issues like variable binding and capture have caused errors in older versions.
\hlint also performs various transformations to apply the hint in different
circumstances. E.g., \eqref{revrev} may be applied to expressions such as \hs{reverse $ reverse xs}, where the \hs{$} is translated away for matching purposes. Another limitation is that \hlint does not perform full name resolution, approximating what set of names a particular identifier may refer to when searching for replacements.

Not an issue of validity but of expressiveness is our treatment of $\bot$: While
to the Haskell developer, exceptions, pattern-match failures, system crashes,
deadlocks and nontermination are very different things, in our semantics of the
language, all of them are modeled as $\bot$, which is unique in every type. So
for the purposes of the HOLCF-Prelude,
\Snippet{head-last}
is a theorem, while no one would want to replace \hs{head []} with
\hs{last (repeat 1)} in their code.

\paragraph{Related Work.}

To formally verify Haskell code there are two main approaches: 1) formalize
functions and their desired properties in a proof assistant and then generate
Haskell code; or 2) translate Haskell code to the language of a proof assistant
and then prove the desired properties.

Generating Haskell code (that is correct by construction) is supported in several proof assistants, for example Isabelle
\cite{isabelle-codegen} and Coq \cite{coqextract}. Such generated code can be
called from normal Haskell code, but must originally be written in the language of the proof assistant, which Haskell programmers may find burdensome.

Translating Haskell code is the approach taken by tools such as Haskabelle
\cite{haskabelle} which produces Isabelle/HOL specifications. Haskell code
can also be manually translated to syntactically similar languages such as Agda
\cite{agda}. Many of these approaches work in the setting of a strict language
but fail to express propositions about laziness, undefinedness and infiniteness.
Results obtained this way still hold in the lazy setting under certain
conditions, as explained in \cite{fastandloose}.

To preserve the precise semantics of Haskell Abel et al.\ \cite{verifyinghaskell} provide a translation
to Agda where functions are wrapped in an abstract evaluation monad.
However, this yields Agda code that does not
immediately resemble the original Haskell code. Our work allows for translating
Haskell code to Isabelle/HOLCF specifications in a semantics preserving manner, without obscuring the relationship to the original code.

\paragraph{Conclusion and Future Work.}

We have presented the HOLCF-Prelude, our formalization of a large part of Haskell's
standard prelude in Isabelle/HOLCF. Applying this formalization to the rewrites
suggested by \hlint allows us to provide certified hints.  At the time of
writing we have certified 143 of the 322 hints in \hlint's database.
%
%
The usefulness of our approach is supported by the flaws found in the database.
Most of the hints which remain unproven refer to types or type classes not modeled in
our formalization (e.g., \hs{Arrow}, \hs{Functor}, \hs{Monad}) or make
statements about things happening at a lower level (e.g., \hs{IO} and
exceptions).

As future work, the same techniques may be applied to certify rewrites that are
automatically applied by the Haskell compiler GHC \cite{ghc-rewriting}.

\enlargethispage{2em}

\bibliographystyle{abbrv}
\bibliography{references}
\end{document}

%% file: snippets.tex
\DefineSnippet{list-induct}{
  \begin{isabelle}%
{\isasymlbrakk}adm\ P{\isacharsemicolon}\ P\ {\isasymbottom}{\isacharsemicolon}\ P\ {\isacharbrackleft}{\isacharbrackright}{\isacharsemicolon}\ {\isasymAnd}x\ xs{\isachardot}\ P\ xs\ {\isasymLongrightarrow}\ P\ {\isacharparenleft}x\ {\isacharcolon}\ xs{\isacharparenright}{\isasymrbrakk}\ {\isasymLongrightarrow}\ P\ xs%
\end{isabelle}
}
\DefineSnippet{reverse-def}{
\isacommand{fixrec}\isamarkupfalse%
\ reverse\ \isakeyword{where}\isanewline
\ \ {\isachardoublequoteopen}reverse{\isasymcdot}{\isacharbrackleft}{\isacharbrackright}\ {\isacharequal}\ {\isacharbrackleft}{\isacharbrackright}{\isachardoublequoteclose}\ {\isacharbar}\isanewline
\ \ {\isachardoublequoteopen}reverse{\isasymcdot}{\isacharparenleft}x{\isacharcolon}xs{\isacharparenright}\ {\isacharequal}\ reverse{\isasymcdot}xs\ {\isacharplus}{\isacharplus}\ {\isacharbrackleft}x{\isacharbrackright}{\isachardoublequoteclose}%
}
\DefineSnippet{reverse-append}{
\isacommand{lemma}\isamarkupfalse%
\ reverse{\isacharunderscore}append\ {\isacharbrackleft}simp{\isacharbrackright}{\isacharcolon}\isanewline
\ \ \isakeyword{assumes}\ {\isachardoublequoteopen}finite{\isacharunderscore}list\ xs{\isachardoublequoteclose}\ \isakeyword{shows}\ {\isachardoublequoteopen}reverse{\isasymcdot}{\isacharparenleft}xs\ {\isacharplus}{\isacharplus}\ ys{\isacharparenright}\ {\isacharequal}\ reverse{\isasymcdot}ys\ {\isacharplus}{\isacharplus}\ reverse{\isasymcdot}xs{\isachardoublequoteclose}\isanewline
\isadelimproof
\ \ %
\endisadelimproof
\isatagproof
\isacommand{using}\isamarkupfalse%
\ assms\ \isacommand{by}\isamarkupfalse%
\ {\isacharparenleft}induction\ xs{\isacharparenright}\ simp{\isacharunderscore}all%
\endisatagproof
{\isafoldproof}%
\isadelimproof
\endisadelimproof
}
\DefineSnippet{finite-list-reverse}{
\isacommand{lemma}\isamarkupfalse%
\ finite{\isacharunderscore}list{\isacharunderscore}reverse\ {\isacharbrackleft}simp{\isacharbrackright}{\isacharcolon}\isanewline
\ \ \isakeyword{assumes}\ {\isachardoublequoteopen}finite{\isacharunderscore}list\ xs{\isachardoublequoteclose}\ \isakeyword{shows}\ {\isachardoublequoteopen}finite{\isacharunderscore}list\ {\isacharparenleft}reverse{\isasymcdot}xs{\isacharparenright}{\isachardoublequoteclose}\isanewline
\isadelimproof
\ \ %
\endisadelimproof
\isatagproof
\isacommand{using}\isamarkupfalse%
\ assms\ \isacommand{by}\isamarkupfalse%
\ {\isacharparenleft}induction\ xs{\isacharparenright}\ simp{\isacharunderscore}all%
\endisatagproof
{\isafoldproof}%
\isadelimproof
\endisadelimproof
}
\DefineSnippet{reverse-reverse}{
\isacommand{lemma}\isamarkupfalse%
\ reverse{\isacharunderscore}reverse{\isacharcolon}\isanewline
\ \ \isakeyword{assumes}\ {\isachardoublequoteopen}finite{\isacharunderscore}list\ xs{\isachardoublequoteclose}\ \isakeyword{shows}\ {\isachardoublequoteopen}reverse{\isasymcdot}{\isacharparenleft}reverse{\isasymcdot}xs{\isacharparenright}\ {\isacharequal}\ xs{\isachardoublequoteclose}\isanewline
\isadelimproof
\ \ %
\endisadelimproof
\isatagproof
\isacommand{using}\isamarkupfalse%
\ assms\ \isacommand{by}\isamarkupfalse%
\ {\isacharparenleft}induction\ xs{\isacharparenright}\ simp{\isacharunderscore}all%
\endisatagproof
{\isafoldproof}%
\isadelimproof
\endisadelimproof
}
\DefineSnippet{reverse-append-below}{
\isacommand{lemma}\isamarkupfalse%
\ reverse{\isacharunderscore}append{\isacharunderscore}below{\isacharcolon}\ {\isachardoublequoteopen}reverse{\isasymcdot}{\isacharparenleft}xs\ {\isacharplus}{\isacharplus}\ ys{\isacharparenright}\ {\isasymsqsubseteq}\ reverse{\isasymcdot}ys\ {\isacharplus}{\isacharplus}\ reverse{\isasymcdot}xs{\isachardoublequoteclose}\isanewline
\isadelimproof
\endisadelimproof
\isatagproof
\isacommand{proof}\isamarkupfalse%
\ {\isacharparenleft}induction\ xs{\isacharparenright}\isanewline
\ \ \isacommand{case}\isamarkupfalse%
\ {\isacharparenleft}Cons\ x\ xs{\isacharparenright}\isanewline
\ \ \isacommand{have}\isamarkupfalse%
\ {\isachardoublequoteopen}reverse{\isasymcdot}{\isacharparenleft}x{\isacharcolon}xs\ {\isacharplus}{\isacharplus}\ ys{\isacharparenright}\ {\isacharequal}\ reverse{\isasymcdot}{\isacharparenleft}xs\ {\isacharplus}{\isacharplus}\ ys{\isacharparenright}\ {\isacharplus}{\isacharplus}\ {\isacharbrackleft}x{\isacharbrackright}{\isachardoublequoteclose}\ \isacommand{by}\isamarkupfalse%
\ simp\isanewline
\ \ \isacommand{also}\isamarkupfalse%
\ \isacommand{have}\isamarkupfalse%
\ {\isachardoublequoteopen}{\isasymdots}\ {\isasymsqsubseteq}\ {\isacharparenleft}reverse{\isasymcdot}ys\ {\isacharplus}{\isacharplus}\ reverse{\isasymcdot}xs{\isacharparenright}\ {\isacharplus}{\isacharplus}\ {\isacharbrackleft}x{\isacharbrackright}{\isachardoublequoteclose}\isanewline
\ \ \ \ \isacommand{by}\isamarkupfalse%
\ {\isacharparenleft}rule\ monofun{\isacharunderscore}cfun{\isacharparenright}{\isacharplus}\ {\isacharparenleft}simp{\isacharunderscore}all\ add{\isacharcolon}\ Cons{\isachardot}IH{\isacharparenright}\isanewline
\ \ \isacommand{finally}\isamarkupfalse%
\ \isacommand{show}\isamarkupfalse%
\ {\isacharquery}case\ \isacommand{by}\isamarkupfalse%
\ simp\isanewline
\isacommand{qed}\isamarkupfalse%
\ simp{\isacharunderscore}all%
\endisatagproof
{\isafoldproof}%
\isadelimproof
\endisadelimproof
}
\DefineSnippet{reverse-reverse-below-goal}{
\isacommand{lemma}\isamarkupfalse%
\ {\isachardoublequoteopen}reverse{\isasymcdot}{\isacharparenleft}reverse{\isasymcdot}x{\isacharparenright}\ {\isasymsqsubseteq}\ x{\isachardoublequoteclose}%
\isadelimproof
\endisadelimproof
\isatagproof
}
\DefineSnippet{reverse-reverse-below}{
\isacommand{lemma}\isamarkupfalse%
\ reverse{\isacharunderscore}reverse{\isacharunderscore}below{\isacharcolon}\ {\isachardoublequoteopen}reverse{\isasymcdot}{\isacharparenleft}reverse{\isasymcdot}xs{\isacharparenright}\ {\isasymsqsubseteq}\ xs{\isachardoublequoteclose}\isanewline
\isadelimproof
\endisadelimproof
\isatagproof
\isacommand{proof}\isamarkupfalse%
\ {\isacharparenleft}induction\ xs{\isacharparenright}\isanewline
\ \ \isacommand{case}\isamarkupfalse%
\ {\isacharparenleft}Cons\ x\ xs{\isacharparenright}\isanewline
\ \ \isacommand{have}\isamarkupfalse%
\ {\isachardoublequoteopen}reverse{\isasymcdot}{\isacharparenleft}reverse{\isasymcdot}{\isacharparenleft}x{\isacharcolon}xs{\isacharparenright}{\isacharparenright}\ {\isacharequal}\ reverse{\isasymcdot}{\isacharparenleft}reverse{\isasymcdot}xs\ {\isacharplus}{\isacharplus}\ {\isacharbrackleft}x{\isacharbrackright}{\isacharparenright}{\isachardoublequoteclose}\ \isacommand{by}\isamarkupfalse%
\ simp\isanewline
\ \ \isacommand{also}\isamarkupfalse%
\ \isacommand{have}\isamarkupfalse%
\ {\isachardoublequoteopen}{\isasymdots}\ {\isasymsqsubseteq}\ reverse{\isasymcdot}{\isacharbrackleft}x{\isacharbrackright}\ {\isacharplus}{\isacharplus}\ reverse{\isasymcdot}{\isacharparenleft}reverse{\isasymcdot}xs{\isacharparenright}{\isachardoublequoteclose}\ \isacommand{by}\isamarkupfalse%
\ {\isacharparenleft}rule\ reverse{\isacharunderscore}append{\isacharunderscore}below{\isacharparenright}\isanewline
\ \ \isacommand{also}\isamarkupfalse%
\ \isacommand{have}\isamarkupfalse%
\ {\isachardoublequoteopen}{\isasymdots}\ {\isacharequal}\ x\ {\isacharcolon}\ reverse{\isasymcdot}{\isacharparenleft}reverse{\isasymcdot}xs{\isacharparenright}{\isachardoublequoteclose}\ \isacommand{by}\isamarkupfalse%
\ simp\isanewline
\ \ \isacommand{also}\isamarkupfalse%
\ \isacommand{have}\isamarkupfalse%
\ {\isachardoublequoteopen}{\isasymdots}\ {\isasymsqsubseteq}\ x\ {\isacharcolon}\ xs{\isachardoublequoteclose}\ \isacommand{by}\isamarkupfalse%
\ {\isacharparenleft}simp\ add{\isacharcolon}\ Cons{\isachardot}IH{\isacharparenright}\isanewline
\ \ \isacommand{finally}\isamarkupfalse%
\ \isacommand{show}\isamarkupfalse%
\ {\isacharquery}case\ \isacommand{{\isachardot}}\isamarkupfalse%
\isanewline
\isacommand{qed}\isamarkupfalse%
\ simp{\isacharunderscore}all%
\endisatagproof
{\isafoldproof}%
\isadelimproof
\endisadelimproof
}
\DefineSnippet{head-last}{
\isacommand{lemma}\isamarkupfalse%
\ {\isachardoublequoteopen}head{\isasymcdot}{\isacharbrackleft}{\isacharbrackright}\ {\isacharequal}\ last{\isasymcdot}{\isacharparenleft}repeat{\isasymcdot}{\isadigit{1}}{\isacharparenright}{\isachardoublequoteclose}%
\isadelimproof
\endisadelimproof
\isatagproof
}